\documentclass[11pt]{article}
\usepackage{amsfonts}
\usepackage{graphics,amssymb,amsthm,amsmath,epsf,rotating}
\usepackage{graphicx}
\usepackage{subfigure}

\usepackage[round]{natbib}

\newtheorem{theorem}{Theorem}[section]

\newtheorem{example}{Example}[section]
\numberwithin{equation}{section}

\allowdisplaybreaks

\begin{document}

\date{}
\title{Small Sample Inference for the Common Coefficient of Variation}
\author{M. R. Kazemi$^{1}$, A. A. Jafari$^2,$\thanks{aajafari@yazd.ac.ir} \\
{\small $^1$Department of Statistics, Fasa University, Fasa, Iran}\\
{\small $^2$Department of Statistics, Yazd University, Yazd, Iran}}

\date{}
\maketitle

\begin{abstract}
This paper utilizes the modified signed log-likelihood ratio method for the problem of inference about the common coefficient of variation in several independent normal populations. This method is applicable for both the problem of hypothesis testing and constructing a confidence interval for this parameter.   Simulation studies show that the coverage probability of this proposed approach is close to the confidence coefficient. Also, its expected length is smaller than expected lengths of other competing approaches. In fact, the proposed approach is very satisfactory regardless of  the number of populations and the different values of the common coefficient of variation even for very small sample size.  Finally, we illustrate the proposed method using two real data sets.
\end{abstract}


\section{Introduction}

In many areas of applied statistics including quality control, chemical experiments, biostatistics, financial analysis and medical research, the coefficient of variation (CV) is commonly used as a measure of dispersion and repeatability of data. It is defined as the ratio of the standard deviation to the mean, and applied to compare relative variability of two or more populations. Here, a critical question is whether their CV’s are the same or not.

 For the first time,
\cite{bennett-76}
considered problem of testing the equality of CV’s by assuming independent normal populations. Then,
a modified version of Bennett’s test by
\cite{sh-su-86},
a likelihood ratio test by
\cite{do-di-83},
an asymptotically chi-square  test and a distribution free squared ranks approach by
\cite{miller-91asymptotic,miller-91use},
some Wald tests by
\cite{gu-ma-96},
an invariant test by
\cite{fe-mi-96},
a family of test statistics based
on Renyi's divergence by
  \cite{pa-pa-00},
 likelihood ratio, Wald and
score tests based on inverse CV's by
 \cite{na-ra-03} and
  a likelihood ratio test based on one-step Newton estimators
  by \cite{ve-jo-07}
 are derived for testing the hypothesis that the CV's of normal populations are equal.
  Recently,
  \cite{fu-ts-98,jafari-10,liu-xu-zh-11,ja-ka-13,kr-le-14,kh-sa-14}
  proposed some tests and performed simulation studies to compare sizes and powers of tests.
  Also, \cite{jafari-15inferences}
  proposed   a test for comparing CV's when the populations are not independent.

If the null hypothesis of equality of CV’s is not rejected, then it may be of interest to estimate the unknown common CV. In practice especially in meta-analysis, we may collect independent samples from different populations with a common CV. For inference about the common CV, there has not yet been a well-developed approach for this purpose: some estimators are presented by
\cite{fe-mi-96},
\cite{ahmed-02},
 \cite{forkman-09},
 and
 \cite{be-ja-08}.
   An approximate confidence interval for the common CV is obtained by \cite{ve-jo-07} based on the likelihood ratio approach. Using Monte Carlo simulation,
 \cite{be-ja-08}
    showed that the coverage probability of this confidence interval is close to the confidence coefficient  when the sample sizes are large.
 Using the concepts of generalized p-value
 \cite{ts-we-89}
 and generalized confidence interval
 \cite{weerahandi-93},
 a generalized approach for inference about this parameter is proposed by
 \cite{tian-05},
  and also, two generalized approaches are presented by
  \cite{be-ja-08}.
   Our simulations studies (Tables \ref{tab.sim1}, \ref{tab.sim2} and \ref{tab.sim3}) indicate that
 there are some cases that  the coverage probabilities of  these three generalized confidence intervals are away from confidence coefficient.  In fact, these approaches are very sensitive to  the common CV parameter. For example, their coverage probabilities are close to one when the common CV is large (i.e. it is equal to 0.3 or 0.35).

In this paper, we are interested in the problem of inference about common CV from different independent normal populations and give a confidence interval for it. This method also is applicable for testing hypothesis about the parameter. For this purpose, we will use the modified signed log-likelihood ratio (MSLR) method introduced by
\cite{barndorff-86,barndorff-91}.
It is a higher order likelihood method and has higher order accuracy even when the sample size is small
\cite{lin-13higher}
and successfully is applied in some  settings, for example:  Ratio of means of two independent log-normal distributions
\cite{wu-ji-wo-su-02};
Comparison of means of log-normal distributions
\cite{gill-04};
Inference on ratio of scale parameters of two independent Weibull distributions
\cite{wu-wo-ng-05};
Approximating the F distribution
\cite{wong-08approx};
Testing the difference of the non-centralities of two non-central t distributions
\cite{ch-le-wo-12};
Common mean of several log-normal distributions
\cite{lin-13higher};
Testing equality normal CVs
\cite{kr-le-14};
Comparing two correlation coefficients
\cite{ka-ja-15modified}.

The remainder of this paper is organized as follows: In Section \ref{sec.infer}, we first review three generalized approaches for constructing confidence interval for the common CV parameter, and then describe the MSLR method for this problem. In Section \ref{sec.sim}, we evaluate the methods with respect to coverage probabilities and expected lengths using Monte Carlo simulation. The methods are illustrated using  two real examples in Section \ref{sec.ex}. Some concluding remarks are given in Section \ref{sec.con}.

\section{Inference about the common CV}
\label{sec.infer}
Let $X_{i1},\dots ,X_{{in}_i}$ ($i{\rm =1},2,..,k$) be a random sample of size $n_i$ from a normal distribution with mean ${\mu }_i>0$ and variance ${\tau }^2{\mu }^2_i,$ where the parameter $\tau >0$ is the common CV. The problem of interest is to test and to construct confidence interval for $\tau $. In this section, we first review the proposed approaches based on generalized inference for this parameter, and then an approach is given for inference about the parameter using MSLR method.

\subsection{ Generalized inferences }

\cite{tian-05}
proposed a generalized confidence interval for the common CV and a generalized p-value for testing  a hypothesis about this parameter.
A generalized pivotal variable for the common CV is considered as
\begin{equation}\label{eq.gv1}
G_1=\frac{\sum^k_{i=1}{\left(n_i-1\right)/R_i}}{\sum^k_{i=1}{\left(n_i-1\right)}},
\end{equation}
where $R_i=\frac{{\bar{x}}_i}{s_i}\sqrt{\frac{U_i}{n_i-1}}-\frac{Z_i}{n_i}$, and ${\bar{x}}_i$ and $s^2_i$ are observed values of ${\bar{X}}_i=\frac{1}{n_i}\sum^{n_i}_{j=1}{X_{ij}}$ and $S^2_i=\frac{1}{n_i-1}\sum^{n_i}_{j=1}{{\left(X_{ij}-{\bar{X}}_i\right)}^2}$, respectively, $U_i$ and $Z_i$ are independent random variables with $U_i\sim {\chi }^2_{(n_i-1)}$ and $Z_i\sim N(0,1)$, $i=1,...,k$.

Also, \cite{be-ja-08}
 proposed a generalized pivotal variable for the common CV as
\begin{equation}\label{eq.gv2}
G_2=\frac{n}{\sum^n_{i=1}{\frac{n_i\sqrt{U_i}}{\sqrt{n_i-1}}\frac{{\bar{x}}_i}{s_i}-\sqrt{n}Z}},
\end{equation}
where $Z\sim N\left(0,1\right)$.  They obtained a generalized pivotal variable by combining this and generalized pivotal variable proposed by
\cite{tian-05}
 as
\begin{equation}\label{eq.gv3}
G_3=\frac{1}{2}G_1+\frac{1}{2}G_2,
\end{equation}

\subsection{MSLR method}

  The log-likelihood function based on the full observations can be written as
\[\ell \left({\boldsymbol \theta }\right)=-n{\rm log}\left(\tau \right)-\sum^k_{i=1}{n_i{\rm log}\left({\mu }_i\right)}-\frac{1}{2{\tau }^2}\sum^k_{i=1}{\sum^{n_i}_{j=1}{{\left(\frac{x_{ij}}{{\mu }_i}-1\right)}^2,}}\]
where ${\boldsymbol \theta }={\left(\tau ,{\mu }_1,\dots ,{\mu }_k\right)}'$ and $n=\sum^k_{i=1}{n_i}$.
Let $\hat{{\boldsymbol \theta }}={\left(\hat{\tau },{\hat{\mu }}_1,{\hat{\mu }}_2,\dots ,{\hat{\mu }}_k\right)}'$ be the maximum likelihood estimator (MLE) of the vector parameter ${\boldsymbol \theta }$. There is not a closed form for the MLE's of the unknown parameters of model. But it could be obtained by using a numerical method like the Newton method.

For fixed value of parameter $\tau $, the constrained maximum likelihood estimators (CMLE) of parameters ${\mu }_i,i=1,\dots ,k,$ are obtained by the following explicit form:
\[{\hat{\mu }}_{i\tau }=\frac{\sqrt{{\bar{X}}^2_i+4{\tau }^2\overline{X^2_i}}-{\bar{X}}_i}{2{\tau }^2},\ \ \ \ \ \ i=1,\dots ,k,\]
where   $\overline{X^2_i}=\frac{1}{n_i}\sum^{n_i}_{j=1}{X^2_{ij}}$.

Now, we use the MSLR method which is the modification of traditional signed log-likelihood ratio (SLR) for inference about $\tau $. The  SLR is defined as
\begin{equation}\label{eq.r}
r\left(\tau \right)={\rm sgn}\left(\hat{\tau }-\tau \right){\left(2(\ell (\hat{\theta })-\ell ({\hat{\theta }}_{\tau }))\right)}^{{1}/{2}},
\end{equation}
where $\hat{\tau }$ is the MLE  of $\tau $, $\hat{{\boldsymbol \theta }}$ is the MLE's of unknown parameters, ${\hat{{\boldsymbol \theta }}}_{\tau }=(\tau ,{\hat{\mu }}_{1\tau },\dots ,{\hat{\mu }}_{k\tau })$ is the vector of CMLE's of unknown parameters for a fixed $\tau $ and ${\rm sgn(.)}$ is the sign function.
Based on Wilks' theorem, it is well known that $r\left(\tau \right)$ is asymptotically standard normal distributed  with error of order $O(n^{-1/2})$ (see \cite{co-hi-74}),
 and therefore, an approximate $100\left(1-\alpha \right)\%$ confidence interval for $\tau $ can be obtained from
\[\left\{\tau :\ \left|r\left(\tau \right)\right|\le Z_{{\alpha }/{2}}\right\},\]
where $Z_{{\alpha }/{2}}$ is the $100\left(1-{\alpha }/{{\rm 2}}\right)\%$th percentile of the standard normal distribution.
{\  \cite{ve-jo-07}
utilized the  likelihood ratio approach and proposed an asymptotic confidence interval for the common CV using Newton one-step estimator. But \cite{be-ja-08} showed that the coverage probability of the confidence interval proposed by
\cite{ve-jo-07}
is smaller than the confidence coefficient when the sample sizes are small. So this approach is not included in our comparison study.}

Generally, \cite{pi-pe-92}
showed the SLR method is not very accurate and some modifications are needed to increase the accuracy of the SLR method. There exist various ways to improve the accuracy of this approximation by adjusting the SLR statistic. For the various ways to improve the accuracy of SLR method, refer to the works of
\cite{barndorff-86,barndorff-91,fr-re-wu-99,Skovgaard-01,di-ma-st-01}.

In this paper, we used the method proposed by
\cite{fr-re-wu-99}
which has the form
\begin{equation}\label{eq.rs}
r^*\left(\tau \right)=r\left(\tau \right)-\frac{1}{r\left(\tau \right)}{\log  \frac{r\left(\tau \right)}{Q\left(\tau \right)}\ },
\end{equation}
where
\[Q(\tau )=\frac{\left| \begin{array}{cc}
{\ell }_{;{\boldsymbol V}}(\hat{{\boldsymbol \theta }})-{\ell }_{;{\boldsymbol V}}({\hat{{\boldsymbol \theta }}}_{\tau }) &
{\ell }_{{\boldsymbol \lambda };{\boldsymbol V}}({\hat{{\boldsymbol \theta }}}_{\tau }) \end{array}
\right|}{\left|{\ell }_{{\boldsymbol \theta };{\boldsymbol V}}(\hat{{\boldsymbol \theta }})\right|}{\left\{\frac{\left|j_{\theta {\theta }'}
(\hat{{\boldsymbol \theta }})\right|}{\left|j_{{\boldsymbol \lambda }{{\boldsymbol \lambda }}^{{\boldsymbol '}}}({\hat{{\boldsymbol \theta }}}_{\tau })\right|}\right\}}^{1/2},\]
and
{\  $j_{{\boldsymbol \theta }{\boldsymbol \theta '}}(\hat{\boldsymbol \theta }) ={\left.\frac{\partial^ 2\ell ({\boldsymbol \theta })}{\partial{\boldsymbol \theta }{\partial{\boldsymbol \theta }}'}\right|}_{{\boldsymbol \theta } =\hat{{\boldsymbol \theta }}}$ and $j_{\boldsymbol \lambda {{\boldsymbol \lambda }}'}({\hat{{\boldsymbol \theta }}}_{\tau }) ={\left.\frac{\partial^2\ell ({\boldsymbol \theta })}{\partial{\boldsymbol \lambda }{\partial{\boldsymbol \lambda }}'}\right|}_{{\boldsymbol \theta } ={\hat{{\boldsymbol \theta }}}_\tau }$}
are the observed information matrix evaluated at $\hat{{\boldsymbol \theta }}$
 and observed nuisance information matrix evaluated at ${\hat{{\boldsymbol \theta }}}_{\tau }$, respectively, and ${\ell }_{;{\boldsymbol V}}({\boldsymbol \theta })$ is the likelihood gradient. Also, the quantity ${\ell }_{{\boldsymbol \theta };{\boldsymbol V}}(\hat{\theta })$ and ${\ell }_{{\boldsymbol \lambda };{\boldsymbol V}}({\hat{{\boldsymbol \theta }}}_{\tau })$ are defined as
\[{\ell }_{{\boldsymbol \theta };{\boldsymbol V}}(\hat{{\boldsymbol \theta }})={\left.\frac{\partial {\ell }_{;{\boldsymbol V}}({\boldsymbol \theta })}{\partial {\boldsymbol \theta }}\right|}_{{\boldsymbol \theta }=\hat{{\boldsymbol \theta }}}\ \ \ \  {\rm and}\ \ \ \ {\ell }_{{\boldsymbol \lambda };{\boldsymbol V}}({\hat{{\boldsymbol \theta }}}_{\tau })={\left.\frac{\partial {\ell }_{;{\boldsymbol V}}({\boldsymbol \theta })}{\partial {\boldsymbol \lambda }}\right|}_{{\boldsymbol \theta }={\hat{{\boldsymbol \theta }}}_{\tau }},\]
where, ${\boldsymbol \lambda }$ is the vector of nuisance parameters. The vector array ${\boldsymbol V}$ is defined as
\[{\boldsymbol V}{\left.=-{\left(\frac{\partial {\boldsymbol R}\left({\boldsymbol X};{\boldsymbol \theta }\right)}{\partial {\boldsymbol X}}\right)}^{-1}\left(\frac{\partial {\boldsymbol R}\left({\boldsymbol X};{\boldsymbol \theta }\right)}{\partial {\boldsymbol \theta }}\right)\right|}_{\hat{{\boldsymbol \theta }}},\]
where ${\boldsymbol R}\left({\boldsymbol X};{\boldsymbol \theta }\right)=\left(R_{11}\left({\boldsymbol X};{\boldsymbol \theta }\right),\dots ,R_{k,n_k}\left({\boldsymbol X};{\boldsymbol \theta }\right)\right)$ is a vector of pivotal quantity.

\begin{theorem}\label{thm.1} (\cite{barndorff-91,fr-re-wu-99})
Generally, $r^*(\tau)$  in \eqref{eq.rs} is asymptotically standard normally distributed with error of order $O(n^{-3/2})$.
\end{theorem}

Based on Theorem \ref{thm.1}, a $100\left(1-\alpha \right)\%$ confidence interval for $\tau $ is given as
\[\left\{\tau :\left|r^*\left(\tau \right)\right|<Z_{\alpha /2}\right\}.\]
Also,
the test statistic $r^*\left(\tau_0 \right)$ can be used for testing the hypotheses $H_0:\ \tau ={\tau }_0$ vs $H_1:\ \tau \ne {\tau }_0$, and the p-value  is given as
\[{\rm p}=2{\rm min}\left\{P\left(Z>r^*\left({\tau }_0\right)\right),P\left(Z<r^*\left({\tau }_0\right)\right)\right\},\]
where $Z$ has a standard normal distribution.

  For our problem in this paper, ${\boldsymbol \lambda }={\left({\mu }_1,\dots ,{\mu }_k\right)}'$ and the details of implementation of $r^*$ are given as follows:

For $i=1,\dots ,k,j=1,..,n_i$, define vector pivotal quantity ${\boldsymbol R}=(R_{11},\dots,
\break
R_{kn_k})'$ with elements $R_{ij}=\frac{x_{ij}-{\mu }_i}{\tau {\mu }_i}$.
The derivative of elements of vector pivotal quantity ${\boldsymbol R}$ with respect to $x_{tj}$ and vector parameter ${\boldsymbol \theta }$ are obtained as
\[\frac{\partial R_{ij}}{{\partial x}_{i'j}}=
\left\{ \begin{array}{lc}
\frac{1}{\tau {\mu }_i} & \ \ i=i' \\
0 & \ \ i\ne i', \end{array}
\right.\ \ \ \ \ \ \
 \frac{\partial R_{ij}}{{\partial\mu }_i'}=\left\{ \begin{array}{lc}
-\frac{x_{ij}}{\tau {\mu }^2_i} & \ \ i=i' \\
0 &  \ \  i\ne i', \end{array}
\right.\ \ \ \ \ \
\frac{\partial R_{ij}}{\partial\tau }=-\frac{x_{ij}-{\mu }_i}{{\tau }^2{\mu }_i}.\]
Therefore, we have
\[{\left(\frac{\partial {\boldsymbol R}\left({\boldsymbol x};{\boldsymbol \theta }\right)}{\partial {\boldsymbol x}}\right)}^{-1}=\tau \left[ \begin{array}{cccc}
{\mu }_1I_{n_1} & {{\boldsymbol 0}}_{n_2} & \cdots  & {{\boldsymbol 0}}_{n_2} \\
{{\boldsymbol 0}}_{n_1} & {\mu }_2I_{n_2} & \cdots  & {{\boldsymbol 0}}_{n_2} \\
\vdots  & \vdots  & \ddots  & \vdots  \\
{{\boldsymbol 0}}_{n_1} & {{\boldsymbol 0}}_{n_2} & \cdots  & {\mu }_kI_{n_k} \end{array}
\right],\]
where ${{\boldsymbol 0}}_{n_i}$ and $I_{n_i}$ are the $n_i\times n_i$ zero and identity matrices, respectively.
Therefore, elements of vector array ${\boldsymbol V}=\left(\boldsymbol V'_1,\dots ,{{\boldsymbol V}}'_{k+1}\right)$ are
\begin{eqnarray*}
&&{{\boldsymbol V}}_1=\left(\frac{x_{11}-{\hat{\mu }}_1}{\hat{\tau }},\dots ,\frac{x_{1n_1}-{\hat{\mu }}_1}{\hat{\tau }},\frac{x_{21}-{\hat{\mu }}_2}{\hat{\tau }},\dots,\right.\\
&&
\hspace{1cm}\left.
 \frac{x_{2n_2}-{\hat{\mu }}_2}{\hat{\tau }},\dots ,\frac{x_{k1}-{\hat{\mu }}_k}{\hat{\tau }},\dots ,\frac{x_{kn_k}-{\hat{\mu }}_k}{\hat{\tau }}\right),\\
&&{{\boldsymbol V}}_2=\left(\frac{x_{11}}{{\hat{\mu }}_1},\dots ,\frac{x_{1n_1}}{{\hat{\mu }}_1},0,\dots ,0\right),\\
&&{{\boldsymbol V}}_3=\left(0,\dots ,0,\frac{x_{21}}{{\hat{\mu }}_2},\dots ,\frac{x_{2n_2}}{{\hat{\mu }}_2},0,\dots ,0\right),\\
&&\ \ \ \ \ \ \vdots\\
&&{{\boldsymbol V}}_{k+1}=\left(0,\dots ,0,0,\dots ,0,\frac{x_{k1}}{{\hat{\mu }}_k},\dots ,\frac{x_{kn_k}}{{\hat{\mu }}_k}\right).
\end{eqnarray*}

The derivative of the log-likelihood function respect to $x_{ij}$ are
$\frac{\partial \ell \left({\boldsymbol \theta }\right)}{{\partial x}_{ij}}= -\frac{x_{ij}-{\mu }_i}{{\mu }^2_i{\tau }^2},
$
$i=1,\dots ,k$, $ j=1,\dots ,n_i.$
The likelihood gradient ${\ell }_{;{\boldsymbol V}}\left({\boldsymbol \theta }\right)$ is obtained as
\[{\ell }_{;{\boldsymbol V}}\left({\boldsymbol \theta }\right)=\left(\sum_j{\frac{\partial \ell \left({\boldsymbol \theta }\right)}{\partial x_j}}v_{1j},\sum_j{\frac{\partial \ell \left({\boldsymbol \theta }\right)}{{\partial x}_j}}v_{2j},\dots ,\sum_j{\frac{\partial \ell \left({\boldsymbol \theta }\right)}{{\partial x}_j}}v_{\left(k+1\right)j}\right).\]
For our problem, this likelihood gradient is obtained as
\begin{align*}
{\ell }_{;{\boldsymbol V}}\left({\boldsymbol \theta }\right)=&\left(\frac{1}{\hat{\tau }{\tau }^2}\sum^k_{i=1}{\sum^{n_i}_{j=1}{\frac{1}{{\mu }^2_i}\left(x_{ij}-{\hat{\mu }}_i\right)\left({\mu }_i-x_{ij}\right)}},\right.
\frac{1}{{\hat{\mu }}_1{{\mu }^2_1\tau }^2}\sum^{n_1}_{j=1}{x_{1j}\left({\mu }_1-x_{1j}\right)},\\
&\hspace{1cm}
\left.\dots ,\frac{1}{{\hat{\mu }}_k{{\mu }^2_k\tau }^2}\sum^{n_k}_{j=1}{x_{kj}\left({\mu }_k-x_{kj}\right)}\right)'.
\end{align*}
Also, the quantity ${\ell }_{{\boldsymbol \theta };{\boldsymbol V}}\left({\boldsymbol \theta }\right)=\frac{\partial {\ell }_{{\boldsymbol \theta };{\boldsymbol V}}\left({\boldsymbol \theta }\right)}{\partial {\boldsymbol \theta }}$ is obtained as
\begin{align*}
{\ell }_{\tau ;{\boldsymbol V}}\left({\boldsymbol \theta }\right)=&\left[\frac{2}{\hat{\tau }{\tau }^3}\sum^k_{i=1}\sum^{n_i}_{j=1}\frac{1}{{\mu }^2_i}\left(x_{ij}-{\hat{\mu }}_i\right)\left(x_{ij}-{\mu }_i\right)\right.,\frac{2}{{\hat{\mu }}_1{{\mu }^2_1\tau }^2}\sum^{n_1}_{j=1}{x_{1j}\left(x_{1j}-{\mu }_1\right)},\\
&\left.\dots ,\frac{2}{{\hat{\mu }}_k{{\mu }^2_k\tau }^2}\sum^{n_k}_{j=1}{x_{kj}\left(x_{kj}-{\mu }_k\right)}\right],\\
{\ell }_{{\mu }_1;{\boldsymbol V}}\left({\boldsymbol \theta }\right)=&\left[\frac{1}{\hat{\tau }{\tau }^2}\sum^{n_1}_{j=1}\left(x_{1j}-{\hat{\mu }}_1\right)\right.\left(\frac{1}{{\mu }^2_1}\right.-\left.\frac{2\left({\mu }_1-x_{1j}\right)}{{\mu }^3_1}\right),
\\
&\frac{1}{{\hat{\mu }}_1{\tau }^2}\sum^{n_1}_{j=1}x_{1j}\left(\frac{1}{{\mu }^2_1}-\frac{2\left({\mu }_1-x_{1j}\right)}{{\mu }^3_1}\right),
 \left. 0,\dots ,0\right],\\
&\vdots \\
{\ell }_{{\mu }_k;{\boldsymbol V}}\left({\boldsymbol \theta }\right)=&\left[\frac{1}{\hat{\tau }{\tau }^2}\sum^{n_k}_{j=1}\left(x_{kj}-{\hat{\mu }}_k\right)\left(\frac{1}{{\mu }^2_k}-\frac{2\left({\mu }_k-x_{kj}\right)}{{\mu }^3_k}\right)\right.
,0,\dots ,0,\\
&
\left.\frac{1}{{\hat{\mu }}_k{\tau }^2}\sum^{n_k}_{j=1}x_{kj}\left(\frac{1}{{\mu }^2_k}-\frac{2\left({\mu }_k-x_{kj}\right)}{{\mu }^3_k}\right)\right].
\end{align*}

We also need to compute the observed information matrix and observed nuisance information matrix. The elements of the observed information matrix is obtained as
\begin{eqnarray*}
&&\ j_{{\mu }_i{\mu }_{i'}}\left({\boldsymbol\theta} \right)=
\left\{ \begin{array}{lc}
-\frac{n_i}{{\mu }^2_i}+\frac{1}{{\tau }^2{\mu }^4_i}\sum^{n_i}_{j=1}{x^2_{ij}+}\frac{2}{{\tau }^2{\mu }^3_i}\sum^{n_i}_{j=1}{x_{ij}\left(\frac{x_{ij}}{{\mu }_i}-1\right)} & \ \ \  i=i' \\
0 & \ \ \ i\ne i', \end{array}
\right.\\
&&j_{\tau \tau }\left({\boldsymbol\theta}\right)=-\frac{\sum^k_{i=1}{n_i}}{{\tau }^2}+\frac{3}{{\tau }^4}\left(\sum^k_{i=1}{\sum^{n_i}_{j=1}{{\left(\frac{x_{ij}}{{\mu }_i}-1\right)}^2}}\right),\\
&&
j_{{\tau ,\mu }_i}\left({\boldsymbol\theta} \right)=\frac{2}{{\tau }^3{\mu }^2_i}\sum^{n_i}_{j=1}{x_{ij}\left(\frac{x_{ij}}{{\mu }_i}-1\right)}.
\end{eqnarray*}
{\  By using the elements $j_{{\mu }_i{\mu }_{i'}}\left({\mathbf \theta }\right)$ for $i,i'=1,\dots ,k$, one can constitute the observed nuisance information matrix.}

\section{Simulation study}
\label{sec.sim}
A simulation study is performed to evaluate the operation of the proposed approach. We performed this with 10,000 replications to compare the coverage probabilities (CP) and expected lengths (EL) of four approaches: the modified signed likelihood ratio (MSLR) method,
generalized pivotal approach in \eqref{eq.gv1} shown by  GV1,
generalized pivotal approach in \eqref{eq.gv2} shown by  GV2, and
generalized pivotal approach in \eqref{eq.gv3} shown by  GV3.

{\  we generate random samples of size $n_i$ from $k=3, 5, 10$ independent normal distributions. We take the true value of model parameter as $\left({\mu }_1,{\mu }_2,{\mu }_3\right)=\left(20,10,10\right)$  for $k=3$,  $({\mu}_1,\dots,{\mu }_5)=(50,40,30,20,10)$ for $k=5$, and
$({\mu}_1,\dots,$ ${\mu }_{10})=(50,40,30,20,10,50,40,30,20,10)$ for $k=10$. The variances of normal populations are obtained such that we have the value of common CV, $\tau $. This value varies in the set $\left\{0.1,0.2,0.3,0.35\right\}$.} For different values of common CV, $\tau $, the coverage probabilities and expected lengths of the MSLR and GV approaches are estimated to construct the confidence interval with the $0.95$ confidence coefficient. The results are given in Tables \ref{tab.sim1}, \ref{tab.sim2} and \ref{tab.sim3}. We can conclude that

\begin{enumerate}
 \item[i.]    The coverage probability of the MSLR method  is  close to the confidence coefficient for all cases. In fact, it is very satisfactory regardless of the number of samples and for all different values of common CV, even for small sample sizes.

  \item[ii.]   The coverage probability of the GV2 is very smaller than the confidence coefficient in most cases.

  \item[iii.]    The coverage probabilities of the GV1 and GV3 are very larger than the confidence coefficient especially when $\tau$ is large (i.e. 0.3 and 0.35). These cases are marked boldface in the tables.

  \item[iv.]    In all cases, the expected length of the MSLR method is shorter than expected lengths of the GV methods, even for the cases that the GV methods act well (i.e. when  their coverage probabilities are close to the confidence coefficient).

  \item[v.] The expected length of the GV1 method is considerably larger than expected lengths of other methods.

  \item[vi.]
The expected lengths of all approaches increase  when the value of $\tau$ increases.   Also, the expected lengths become smaller when the sample sizes increase.

\end{enumerate}

Since, the MSLR method is the only approach that controls the correct confidence coefficient and has the shorter interval length with respect to the other competing approaches in all cases, we recommend researchers use the MSLR method for practical applications when the random samples are normal.

  To compare robustness of the MSLR and  GV approaches, a similar simulation study is performed by considering the Weibull distribution with shape parameter $\alpha $ and scale parameter $\beta$ as the following probability density function:
$$
f\left(x\right)=\frac{\alpha }{\beta }{(\frac{x}{\beta })}^{\alpha -1}{\exp  (-(\frac{x}{\beta })^{\alpha })},\ \  x>0.
$$
The random samples are generated from $k$ Weibull distributions, and the parameters are chosen such that a common CV, $\tau$, holds.
 We take the true value of model parameter as $(\beta_1,\beta_2,\beta_3)=(20,10,10)$  for $k=3$,
 $(\beta_1,\dots,\beta_5)=(50,40,30,20,10)$ for $k=5$, and
$({\beta}_1,\dots,{\beta }_{10})=(50,40,30,20,10,50,40,30,20, $ $10)$ for $k=10$ where $\beta_i$ is the scale parameter of $i$th Weibull distribution.
 The results are given in Tables \ref{tab.sim4}, \ref{tab.sim5} and \ref{tab.sim6}.  We can conclude that
 the coverage probability of the MSLR method  is  close to the confidence coefficient when  $\tau$ is large (i.e. 0.3 and 0.35) and is smaller than the confidence coefficient for other cases. Other results are similar to those reported in normal case.

\begin{table}
\begin{center}
\caption{  Empirical coverage probabilities and expected lengths of two-sided confidence intervals for the parameter of common CV under normal distribution for $k=3$.}\label{tab.sim1}

\begin{tabular}{|c|c|cccc|cccc|} \hline
 &  & \multicolumn{4}{|c|}{$\tau =0.1$} & \multicolumn{4}{|c|}{$\tau =0.2$} \\ \hline
$n_1,n_2,n_3$ &  & MSLR & GV1 & GV2 & GV3 & MSLR & GV1 & GV2 & GV3 \\ \hline

4,4,4 & CP & 0.950 & 0.954~ & 0.906~ & 0.964 & 0.950 & \textbf{0.972} & 0.918 & \textbf{0.971} \\
 & EL & 0.108 & ~0.208 & 0.107~ & 0.147 & 0.222 & 0.522 & 0.224 & 0.343 \\ \hline

4,5,6 & CP & 0.943 & 0.956 & 0.918 & 0.956 & 0.953 & 0.961 & 0.913 & 0.958 \\
 & EL & 0.091 & 0.146 & 0.090 & 0.111 & 0.186 & 0.343 & 0.186 & 0.245 \\ \hline

6,5,4 & CP & 0.936 & 0.954 & 0.918 & 0.956 & 0.933 & 0.963 & 0.921 & 0.949 \\
 & EL & 0.091 & 0.146 & 0.090 & 0.110 & 0.189 & 0.344 & 0.187 & 0.247 \\ \hline

5,5,10 & CP & 0.942 & 0.956 & 0.927 & 0.954 & 0.940 & 0.957 & 0.923 & 0.958 \\
 & EL & 0.090 & 0.103 & 0.073 & 0.084 & 0.189 & 0.233 & 0.152 & 0.181 \\ \hline

10,5,5 & CP & 0.938 & 0.958 & 0.927 & 0.954 & 0.954 & 0.953 & 0.922 & 0.960 \\
 & EL & 0.074 & 0.103 & 0.073 & 0.084 & 0.153 & 0.233 & 0.152 & 0.180 \\ \hline

4,5,20 & CP & 0.944 & 0.953 & 0.923 & 0.956 & 0.957 & 0.960 & 0.927 & 0.955 \\
 & EL & 0.058 & 0.077 & 0.059 & 0.064 & 0.121 & 0.173 & 0.121 & 0.139 \\ \hline

20,5,4 & CP & 0.942 & 0.955 & 0.925 & 0.950 & 0.954 & 0.963 & 0.929 & 0.953 \\
 & EL & 0.058 & 0.077 & 0.059 & 0.064 & 0.121 & 0.174 & 0.121 & 0.140 \\ \hline

7,7,7 & CP & 0.954 & 0.956 & 0.929 & 0.949 & 0.938 & 0.956 & 0.926 & 0.954 \\
 & EL & 0.072 & 0.094 & 0.071 & 0.079 & 0.149 & 0.206 & 0.146 & 0.166 \\ \hline

7,8,9 & CP & 0.954 & 0.958 & 0.931 & 0.951 & 0.944 & 0.954 & 0.933 & 0.954 \\
 & EL & 0.066 & 0.082 & 0.065 & 0.071 & 0.136 & 0.179 & 0.134 & 0.149 \\ \hline

 &  & \multicolumn{4}{|c|}{$\tau =0.3$} & \multicolumn{4}{|c|}{$\tau =0.35$} \\ \hline

4,4,4 & CP & 0.949 & \textbf{0.999} & 0.915 & \textbf{0.990} & 0.951 & \textbf{0.999} & 0.919 & \textbf{0.983} \\
 & EL & 0.354 & 1.160 & 0.358 & 0.680 & 0.432 & 1.774 & 0.436 & 0.593 \\ \hline

4,5,6 & CP & 0.953 & \textbf{0.983} & 0.920 & \textbf{0.978} & 0.955 & \textbf{0.997} & 0.926 & \textbf{0.983} \\
 & EL & 0.295 & 0.687 & 0.298 & 0.437 & 0.358 & 0.985 & 0.361 & 0.593 \\ \hline

6,5,4 & CP & 0.938 & \textbf{0.982} & 0.922 & \textbf{0.979} & 0.950 & \textbf{0.997} & 0.919 & \textbf{0.987} \\
 & EL & 0.302 & 0.685 & 0.297 & 0.453 & 0.364 & 0.976 & 0.361 & 0.598 \\ \hline

5,5,10 & CP & 0.945 & 0.963 & 0.923 & 0.960 & 0.948 & \textbf{0.979} & 0.936 & \textbf{0.967} \\
 & EL & 0.302 & 0.432 & 0.241 & 0.308 & 0.369 & 0.588 & 0.290 & 0.396 \\ \hline

10,5,5 & CP & 0.955 & \textbf{0.969} & 0.925 & \textbf{0.965} & 0.946 & \textbf{0.975} & 0.923 & \textbf{0.973} \\
 & EL & 0.245 & 0.428 & 0.240 & 0.312 & 0.294 & 0.585 & 0.291 & 0.404 \\ \hline

4,5,20 & CP & 0.936 & \textbf{0.974} & 0.927 & 0.961 & 0.952 & \textbf{0.983} & 0.930 & \textbf{0.973} \\
 & EL & 0.191 & 0.323 & 0.190 & 0.236 & 0.227 & 0.441 & 0.229 & 0.308 \\ \hline

20,5,4 & CP & 0.944 & \textbf{0.973} & 0.927 & \textbf{0.970} & 0.951 & \textbf{0.984} & 0.930 & \textbf{0.978} \\
 & EL & 0.190 & 0.323 & 0.190 & 0.237 & 0.227 & 0.445 & 0.229 & 0.308 \\ \hline

7,7,7 & CP & 0.954 & 0.952 & 0.926 & 0.963 & 0.949 & 0.956 & 0.928 & 0.945 \\
 & EL & 0.237 & 0.368 & 0.232 & 0.278 & 0.285 & 0.482 & 0.280 & 0.349 \\ \hline

7,8,9 & CP & 0.945 & 0.959 & 0.934 & 0.954 & 0.944 & 0.955 & 0.932 & 0.939 \\
 & EL & 0.215 & 0.308 & 0.211 & 0.240 & 0.260 & 0.402 & 0.256 & 0.304 \\ \hline
\end{tabular}

\end{center}
\end{table}

\begin{table}
\begin{center}
\caption{  Empirical coverage probabilities and expected lengths of two-sided confidence intervals for the parameter of common CV under normal distribution for $k=5$.}\label{tab.sim2}

\begin{tabular}{|c|c|cccc|cccc|} \hline
 &  & \multicolumn{4}{|c|}{$\tau =0.1$} & \multicolumn{4}{|c|}{$\tau =0.2$} \\ \hline
 $n_1,\dots ,n_5$ &  & MSLR & GV1 & GV2 & GV3 & MSLR & GV1 & GV2 & GV3 \\ \hline

4,4,4,4,4 & CP & 0.947 & 0.934 & 0.869 & \textbf{0.970} & 0.948 & \textbf{0.968} & 0.870 & \textbf{0.984} \\
 & EL & 0.079 & 0.172 & 0.076 & 0.114 & 0.161 & 0.453 & 0.157 & 0.279 \\ \hline

4,4,5,5,6 & CP & 0.956 & 0.938 & 0.883 & 0.962 & 0.958 & 0.947 & 0.883 & \textbf{0.969} \\
 & EL & 0.069 & 0.125 & 0.067 & 0.089 & 0.141 & 0.312 & 0.139 & 0.204 \\ \hline

6,5,5,4,4 & CP & 0.951 & 0.935 & 0.885 & 0.960 & 0.950 & 0.950 & 0.881 & \textbf{0.967} \\
 & EL & 0.069 & 0.126 & 0.068 & 0.089 & 0.141 & 0.309 & 0.138 & 0.205 \\ \hline

5,5,5,5,10 & CP & 0.938 & 0.937 & 0.891 & 0.957 & 0.944 & 0.943 & 0.898 & 0.958 \\
 & EL & 0.059 & 0.092 & 0.058 & 0.070 & 0.121 & 0.217 & 0.120 & 0.155 \\ \hline

10,5,5,5,5 & CP & 0.941 & 0.938 & 0.894 & 0.953 & 0.946 & 0.942 & 0.892 & 0.957 \\
 & EL & 0.060 & 0.092 & 0.059 & 0.070 & 0.122 & 0.217 & 0.120 & 0.155 \\ \hline

4,4,5,5,20 & CP & 0.950 & 0.942 & 0.888 & 0.953 & 0.951 & 0.952 & 0.897 & 0.960 \\
 & EL & 0.051 & 0.078 & 0.051 & 0.060 & 0.105 & 0.187 & 0.105 & 0.135 \\ \hline

20,5,5,4,4 & CP & 0.949 & 0.946 & 0.897 & 0.960 & 0.949 & 0.953 & 0.896 & \textbf{0.966} \\
 & EL & 0.051 & 0.078 & 0.051 & 0.060 & 0.105 & 0.187 & 0.105 & 0.134 \\ \hline

7,7,7,7,7 & CP & 0.950 & 0.939 & 0.902 & 0.955 & 0.957 & 0.939 & 0.907 & 0.953 \\
 & EL & 0.054 & 0.074 & 0.053 & 0.060 & 0.110 & 0.165 & 0.108 & 0.127 \\ \hline

7,7,8,8,9 & CP & 0.954 & 0.939 & 0.912 & 0.952 & 0.959 & 0.941 & 0.914 & 0.951 \\
 & EL & 0.050 & 0.066 & 0.049 & 0.055 & 0.103 & 0.145 & 0.101 & 0.115 \\ \hline

 &  & \multicolumn{4}{|c|}{$\tau =0.3$} & \multicolumn{4}{|c|}{$\tau =0.35$} \\ \hline

4,4,4,4,4 & CP & 0.944 & \textbf{0.999} & 0.873 & \textbf{0.997} & 0.945 & \textbf{1.000} & 0.877 & \textbf{0.998} \\
 & EL & 0.254 & 1.160 & 0.246 & 0.635 & 0.305 & 1.868 & 0.296 & 0.976 \\ \hline

4,4,5,5,6 & CP & 0.957 & \textbf{0.992} & 0.887 & \textbf{0.988} & 0.951 & \textbf{1.000} & 0.890 & \textbf{0.995} \\
 & EL & 0.221 & 0.678 & 0.218 & 0.401 & 0.265 & 1.051 & 0.262 & 0.579 \\ \hline

6,5,5,4,4 & CP & 0.952 & \textbf{0.993} & 0.884 & \textbf{0.991} & 0.950 & \textbf{1.000} & 0.890 & \textbf{0.995} \\
 & EL & 0.221 & 0.681 & 0.218 & 0.403 & 0.265 & 1.032 & 0.261 & 0.582 \\ \hline

5,5,5,5,10 & CP & 0.940 & 0.966 & 0.900 & \textbf{0.972} & 0.941 & \textbf{0.986} & 0.895 & \textbf{0.985} \\
 & EL & 0.190 & 0.425 & 0.187 & 0.280 & 0.228 & 0.618 & 0.225 & 0.379 \\ \hline

10,5,5,5,5 & CP & 0.953 & 0.964 & 0.898 & \textbf{0.971} & 0.941 & \textbf{0.988} & 0.896 & \textbf{0.982} \\
 & EL & 0.192 & 0.430 & 0.188 & 0.280 & 0.231 & 0.614 & 0.225 & 0.379 \\ \hline

4,4,5,5,20 & CP & 0.949 & \textbf{0.982} & 0.902 & \textbf{0.979} & 0.950 & \textbf{0.994} & 0.895 & \textbf{0.988} \\
 & EL & 0.164 & 0.386 & 0.164 & 0.250 & 0.197 & 0.576 & 0.196 & 0.346 \\ \hline

20,5,5,4,4 & CP & 0.952 & \textbf{0.982} & 0.898 & \textbf{0.982} & 0.949 & \textbf{0.994} & 0.899 & \textbf{0.988} \\
 & EL & 0.164 & 0.385 & 0.164 & 0.247 & 0.197 & 0.576 & 0.197 & 0.340 \\ \hline

7,7,7,7,7 & CP & 0.955 & 0.938 & 0.910 & 0.960 & 0.954 & 0.946 & 0.907 & 0.963 \\
 & EL & 0.173 & 0.299 & 0.170 & 0.214 & 0.207 & 0.398 & 0.203 & 0.272 \\ \hline

7,7,8,8,9 & CP & 0.963 & 0.942 & 0.914 & 0.951 & 0.961 & 0.942 & 0.917 & 0.950 \\
 & EL & 0.161 & 0.257 & 0.159 & 0.190 & 0.194 & 0.337 & 0.191 & 0.238 \\ \hline
\end{tabular}

\end{center}
\end{table}

\begin{table}
\begin{center}
\caption{  Empirical coverage probabilities and expected lengths of two-sided confidence intervals for the parameter of common CV under normal distribution for $k=10$.}\label{tab.sim3}

\begin{tabular}{|c|c|cccc|cccc|} \hline
 &  & \multicolumn{4}{|c|}{$\tau =0.1$} & \multicolumn{4}{|c|}{$\tau =0.2$} \\ \hline
$n_1,\dots ,n_{10}$ &  & MSLR & GV1 & GV2 & GV3 & MSLR & GV1 & GV2 & GV3 \\ \hline

4,4,4,4,4,4,4,4,4,4 & CP & 0.941 & 0.866 & 0.753 & 0.970 & 0.944 & \textbf{0.976} & 0.752 & \textbf{0.995} \\
 & EL & 0.053 & 0.132 & 0.051 & 0.083 & 0.107 & 0.390 & 0.104 & 0.224 \\ \hline

4,4,5,5,6,4,4,5,5,6 & CP & 0.946 & 0.883 & 0.781 & 0.965 & 0.946 & 0.910 & 0.785 & \textbf{0.976} \\
 & EL & 0.047 & 0.093 & 0.046 & 0.063 & 0.095 & 0.248 & 0.093 & 0.154 \\ \hline

6,5,5,4,4,6,5,5,4,4 & CP & 0.942 & 0.880 & 0.788 & 0.963 & 0.944 & 0.910 & 0.793 & \textbf{0.977} \\
 & EL & 0.047 & 0.094 & 0.046 & 0.063 & 0.095 & 0.247 & 0.093 & 0.154 \\ \hline

5,5,5,5,10,5,5,5,5,10 & CP & 0.942 & 0.889 & 0.818 & 0.954 & 0.944 & 0.890 & 0.826 & 0.966 \\
 & EL & 0.040 & 0.067 & 0.040 & 0.049 & 0.083 & 0.165 & 0.082 & 0.112 \\ \hline

10,5,5,5,5,10,5,5,5,5 & CP & 0.943 & 0.891 & 0.822 & 0.962 & 0.948 & 0.893 & 0.826 & 0.965 \\
 & EL & 0.040 & 0.067 & 0.040 & 0.049 & 0.083 & 0.165 & 0.082 & 0.112 \\ \hline

4,4,5,5,20,4,4,5,5,20 & CP & 0.945 & 0.898 & 0.818 & 0.960 & 0.944 & 0.925 & 0.830 & \textbf{0.972} \\
 & EL & 0.035 & 0.057 & 0.036 & 0.043 & 0.072 & 0.146 & 0.073 & 0.099 \\ \hline

20,5,5,4,4,20,5,5,4,4 & CP & 0.940 & 0.903 & 0.819 & 0.962 & 0.948 & 0.921 & 0.829 & 0.967 \\
 & EL & 0.035 & 0.057 & 0.036 & 0.043 & 0.072 & 0.146 & 0.073 & 0.099 \\ \hline

7,7,7,7,7,7,7,7,7,7 & CP & 0.942 & 0.899 & 0.846 & 0.954 & 0.947 & 0.892 & 0.852 & 0.955 \\
 & EL & 0.037 & 0.053 & 0.036 & 0.042 & 0.075 & 0.120 & 0.074 & 0.089 \\ \hline

7,7,8,8,9,7,7,8,8,9 & CP & 0.944 & 0.902 & 0.854 & 0.953 & 0.946 & 0.910 & 0.863 & 0.956 \\
 & EL & 0.034 & 0.047 & 0.034 & 0.038 & 0.071 & 0.105 & 0.070 & 0.081 \\ \hline

 &  & \multicolumn{4}{|c|}{$\tau =0.3$} & \multicolumn{4}{|c|}{$\tau =0.35$} \\ \hline

4,4,4,4,4,4,4,4,4,4 & CP & 0.945 & \textbf{1.000} & 0.759 & \textbf{1.000} & 0.946 & \textbf{1.000} & 0.771 & \textbf{1.000} \\
 & EL & 0.167 & 1.235 & 0.161 & 0.630 & 0.199 & 1.964 & 0.192 & 0.991 \\ \hline

4,4,5,5,6,4,4,5,5,6 & CP & 0.950 & \textbf{0.999} & 0.799 & \textbf{0.999} & 0.948 & \textbf{1.000} & 0.800 & \textbf{0.999} \\
 & EL & 0.148 & 0.637 & 0.145 & 0.350 & 0.177 & 1.046 & 0.172 & 0.544 \\ \hline

6,5,5,4,4,6,5,5,4,4 & CP & 0.945 & \textbf{0.999} & 0.796 & \textbf{0.998} & 0.951 & \textbf{1.000} & 0.797 & \textbf{0.998} \\
 & EL & 0.148 & 0.633 & 0.145 & 0.348 & 0.177 & 1.056 & 0.172 & 0.549 \\ \hline

5,5,5,5,10,5,5,5,5,10 & CP & 0.943 & 0.953 & 0.826 & \textbf{0.983} & 0.950 & \textbf{0.991} & 0.838 & \textbf{0.995} \\
 & EL & 0.129 & 0.356 & 0.127 & 0.218 & 0.154 & 0.558 & 0.152 & 0.317 \\ \hline

10,5,5,5,5,10,5,5,5,5 & CP & 0.946 & 0.955 & 0.829 & \textbf{0.983} & 0.946 & \textbf{0.993} & 0.834 & \textbf{0.996} \\
 & EL & 0.129 & 0.357 & 0.127 & 0.218 & 0.154 & 0.561 & 0.152 & 0.318 \\ \hline

4,4,5,5,20,4,4,5,5,20 & CP & 0.947 & \textbf{0.991} & 0.827 & \textbf{0.992} & 0.950 & \textbf{0.999} & 0.836 & \textbf{0.996} \\
 & EL & 0.111 & 0.351 & 0.113 & 0.205 & 0.133 & 0.560 & 0.134 & 0.304 \\ \hline

20,5,5,4,4,20,5,5,4,4 & CP & 0.949 & \textbf{0.992} & 0.836 & \textbf{0.995} & 0.945 & \textbf{0.999} & 0.840 & \textbf{0.997} \\
 & EL & 0.111 & 0.352 & 0.113 & 0.206 & 0.134 & 0.566 & 0.134 & 0.306 \\ \hline

7,7,7,7,7,7,7,7,7,7 & CP & 0.945 & 0.891 & 0.852 & 0.954 & 0.948 & 0.893 & 0.863 & 0.955 \\
 & EL & 0.118 & 0.225 & 0.115 & 0.154 & 0.141 & 0.312 & 0.138 & 0.202 \\ \hline

7,7,8,8,9,7,7,8,8,9 & CP & 0.939 & 0.894 & 0.873 & 0.955 & 0.946 & 0.891 & 0.875 & 0.953 \\
 & EL & 0.108 & 0.191 & 0.109 & 0.136 & 0.131 & 0.257 & 0.130 & 0.174 \\ \hline
\end{tabular}

\end{center}
\end{table}

\begin{table}
\begin{center}
\caption{  Empirical coverage probabilities and expected lengths of two-sided confidence intervals for the parameter of common CV under Weibull distribution for $k=3$.}\label{tab.sim4}

\begin{tabular}{|c|c|cccc|cccc|} \hline
 &  & \multicolumn{4}{|c|}{$\tau =0.1$} & \multicolumn{4}{|c|}{$\tau =0.2$} \\ \hline
$n_1,\dots ,n_{10}$ &  & MSLR & GV1 & GV2 & GV3 & MSLR & GV1 & GV2 & GV3 \\ \hline

4,4,4 & CP & 0.900 & 0.939 & 0.876 & 0.948 & 0.932 & 0.964 & 0.895 & 0.962 \\
 & EL & 0.110 & 0.211 & 0.105 & 0.146 & 0.228 & 0.536 & 0.224 & 0.351 \\ \hline

4,5,6 & CP & 0.894 & 0.933 & 0.884 & 0.937 & 0.927 & 0.950 & 0.911 & 0.951 \\
 & EL & 0.092 & 0.147 & 0.088 & 0.110 & 0.190 & 0.353 & 0.187 & 0.250 \\ \hline

6,5,4 & CP & 0.899 & 0.934 & 0.881 & 0.934 & 0.934 & 0.950 & 0.898 & 0.950 \\
 & EL & 0.092 & 0.147 & 0.088 & 0.110 & 0.190 & 0.351 & 0.186 & 0.249 \\ \hline

5,5,10 & CP & 0.895 & 0.923 & 0.879 & 0.920 & 0.935 & 0.942 & 0.908 & 0.941 \\
 & EL & 0.074 & 0.104 & 0.073 & 0.083 & 0.154 & 0.237 & 0.152 & 0.182 \\ \hline

10,5,5 & CP & 0.895 & 0.927 & 0.887 & 0.925 & 0.935 & 0.941 & 0.908 & 0.940 \\
 & EL & 0.074 & 0.104 & 0.072 & 0.083 & 0.154 & 0.236 & 0.152 & 0.182 \\ \hline

4,5,20 & CP & 0.890 & 0.922 & 0.882 & 0.920 & 0.926 & 0.946 & 0.906 & 0.938 \\
 & EL & 0.058 & 0.078 & 0.058 & 0.064 & 0.120 & 0.176 & 0.121 & 0.139 \\ \hline

20,5,4 & CP & 0.888 & 0.922 & 0.887 & 0.921 & 0.929 & 0.948 & 0.912 & 0.943 \\
 & EL & 0.058 & 0.078 & 0.058 & 0.064 & 0.121 & 0.176 & 0.121 & 0.139 \\ \hline

7,7,7 & CP & 0.894 & 0.921 & 0.885 & 0.916 & 0.932 & 0.942 & 0.911 & 0.937 \\
 & EL & 0.072 & 0.095 & 0.070 & 0.078 & 0.149 & 0.210 & 0.147 & 0.168 \\ \hline

7,8,9 & CP & 0.887 & 0.920 & 0.889 & 0.919 & 0.933 & 0.942 & 0.915 & 0.936 \\
 & EL & 0.066 & 0.083 & 0.064 & 0.070 & 0.136 & 0.181 & 0.134 & 0.149 \\ \hline

 &  & \multicolumn{4}{|c|}{$\tau =0.3$} & \multicolumn{4}{|c|}{$\tau =0.35$} \\ \hline

4,4,4 & CP & 0.962 & \textbf{0.999} & 0.913 & \textbf{0.993} & 0.965 & \textbf{0.999} & 0.927 & \textbf{0.994} \\
 & EL & 0.362 & 1.139 & 0.362 & 0.681 & 0.435 & 1.672 & 0.442 & 0.946 \\ \hline

4,5,6 & CP & 0.956 & \textbf{0.986} & 0.925 & \textbf{0.978} & 0.968 & \textbf{0.998} & 0.934 & \textbf{0.988} \\
 & EL & 0.300 & 0.674 & 0.298 & 0.445 & 0.360 & 0.907 & 0.363 & 0.577 \\ \hline

6,5,4 & CP & 0.961 & \textbf{0.984} & 0.926 & \textbf{0.976} & 0.969 & \textbf{0.999} & 0.935 & \textbf{0.990} \\
 & EL & 0.301 & 0.677 & 0.301 & 0.448 & 0.362 & 0.915 & 0.364 & 0.581 \\ \hline

5,5,10 & CP & 0.956 & \textbf{0.970} & 0.935 & 0.963 & 0.964 & \textbf{0.984} & 0.946 & \textbf{0.977} \\
 & EL & 0.243 & 0.428 & 0.243 & 0.309 & 0.291 & 0.559 & 0.293 & 0.389 \\ \hline

10,5,5 & CP & 0.960 & \textbf{0.972} & 0.935 & 0.966 & 0.964 & \textbf{0.984} & 0.944 & \textbf{0.972} \\
 & EL & 0.243 & 0.428 & 0.243 & 0.309 & 0.292 & 0.559 & 0.294 & 0.390 \\ \hline

4,5,20 & CP & 0.958 & \textbf{0.977} & 0.941 & 0.969 & 0.969 & \textbf{0.989} & 0.949 & \textbf{0.979} \\
 & EL & 0.190 & 0.320 & 0.192 & 0.236 & 0.227 & 0.419 & 0.230 & 0.296 \\ \hline

20,5,4 & CP & 0.959 & \textbf{0.977} & 0.932 & 0.966 & 0.971 & \textbf{0.987} & 0.949 & \textbf{0.978} \\
 & EL & 0.189 & 0.321 & 0.191 & 0.236 & 0.228 & 0.419 & 0.230 & 0.296 \\ \hline

7,7,7 & CP & 0.963 & 0.959 & 0.939 & 0.953 & 0.966 & \textbf{0.972} & 0.949 & 0.965 \\
 & EL & 0.234 & 0.366 & 0.233 & 0.277 & 0.281 & 0.471 & 0.282 & 0.344 \\ \hline

7,8,9 & CP & 0.963 & 0.960 & 0.942 & 0.956 & 0.970 & \textbf{0.970} & 0.952 & 0.962 \\
 & EL & 0.214 & 0.310 & 0.214 & 0.243 & 0.257 & 0.390 & 0.256 & 0.296 \\ \hline
\end{tabular}

\end{center}
\end{table}

\begin{table}
\begin{center}
\caption{  Empirical coverage probabilities and expected lengths of two-sided confidence intervals for the parameter of common CV under Weibull distribution for $k=5$.}\label{tab.sim5}

\begin{tabular}{|c|c|cccc|cccc|} \hline
 &  & \multicolumn{4}{|c|}{$\tau =0.1$} & \multicolumn{4}{|c|}{$\tau =0.2$} \\ \hline
$n_1,\dots ,n_{10}$ &  & MSLR & GV1 & GV2 & GV3 & MSLR & GV1 & GV2 & GV3 \\ \hline

4,4,4,4,4 & CP & 0.894 & 0.912 & 0.821 & 0.951 & 0.928 & 0.959 & 0.852 & \textbf{0.976} \\
 & EL & 0.079 & 0.175 & 0.074 & 0.114 & 0.164 & 0.469 & 0.157 & 0.285 \\ \hline

4,4,5,5,6 & CP & 0.891 & 0.915 & 0.827 & 0.943 & 0.928 & 0.932 & 0.866 & 0.961 \\
 & EL & 0.069 & 0.128 & 0.066 & 0.089 & 0.143 & 0.319 & 0.139 & 0.209 \\ \hline

6,5,5,4,4 & CP & 0.897 & 0.918 & 0.833 & 0.944 & 0.934 & 0.934 & 0.861 & 0.959 \\
 & EL & 0.069 & 0.127 & 0.066 & 0.089 & 0.142 & 0.318 & 0.138 & 0.208 \\ \hline

5,5,5,5,10 & CP & 0.894 & 0.913 & 0.841 & 0.933 & 0.931 & 0.926 & 0.880 & 0.949 \\
 & EL & 0.059 & 0.093 & 0.057 & 0.070 & 0.123 & 0.222 & 0.120 & 0.157 \\ \hline

10,5,5,5,5 & CP & 0.885 & 0.916 & 0.834 & 0.932 & 0.929 & 0.928 & 0.878 & 0.948 \\
 & EL & 0.059 & 0.093 & 0.057 & 0.070 & 0.122 & 0.221 & 0.120 & 0.156 \\ \hline

4,4,5,5,20 & CP & 0.884 & 0.917 & 0.846 & 0.930 & 0.934 & 0.942 & 0.883 & 0.954 \\
 & EL & 0.051 & 0.079 & 0.051 & 0.060 & 0.105 & 0.190 & 0.105 & 0.136 \\ \hline

20,5,5,4,4 & CP & 0.892 & 0.910 & 0.834 & 0.924 & 0.929 & 0.938 & 0.875 & 0.955 \\
 & EL & 0.051 & 0.079 & 0.051 & 0.060 & 0.105 & 0.191 & 0.105 & 0.136 \\ \hline

7,7,7,7,7 & CP & 0.882 & 0.913 & 0.848 & 0.921 & 0.931 & 0.929 & 0.894 & 0.944 \\
 & EL & 0.054 & 0.075 & 0.052 & 0.059 & 0.111 & 0.166 & 0.108 & 0.127 \\ \hline

7,7,8,8,9 & CP & 0.885 & 0.915 & 0.855 & 0.922 & 0.931 & 0.922 & 0.895 & 0.941 \\
 & EL & 0.050 & 0.067 & 0.049 & 0.054 & 0.104 & 0.148 & 0.102 & 0.116 \\ \hline

 &  & \multicolumn{4}{|c|}{$\tau =0.3$} & \multicolumn{4}{|c|}{$\tau =0.35$} \\ \hline

4,4,4,4,4 & CP & 0.960 & \textbf{1.000} & 0.885 & \textbf{0.997} & 0.969 & \textbf{1.000} & 0.886 & \textbf{0.999} \\
 & EL & 0.254 & 1.131 & 0.249 & 0.619 & 0.300 & 1.733 & 0.298 & 0.907 \\ \hline

4,4,5,5,6 & CP & 0.959 & \textbf{0.993} & 0.897 & \textbf{0.991} & 0.970 & \textbf{1.000} & 0.908 & \textbf{0.997} \\
 & EL & 0.222 & 0.663 & 0.219 & 0.399 & 0.264 & 0.968 & 0.265 & 0.553 \\ \hline

6,5,5,4,4 & CP & 0.957 & \textbf{0.993} & 0.898 & \textbf{0.990} & 0.968 & \textbf{1.000} & 0.910 & \textbf{0.997} \\
 & EL & 0.222 & 0.666 & 0.220 & 0.401 & 0.265 & 0.964 & 0.264 & 0.550 \\ \hline

5,5,5,5,10 & CP & 0.958 & 0.966 & 0.911 & \textbf{0.975} & 0.970 & \textbf{0.990} & 0.923 & \textbf{0.989} \\
 & EL & 0.191 & 0.426 & 0.189 & 0.280 & 0.228 & 0.577 & 0.227 & 0.364 \\ \hline

10,5,5,5,5 & CP & 0.960 & 0.962 & 0.902 & \textbf{0.972} & 0.970 & \textbf{0.987} & 0.923 & \textbf{0.984} \\
 & EL & 0.191 & 0.424 & 0.189 & 0.279 & 0.228 & 0.579 & 0.228 & 0.365 \\ \hline

4,4,5,5,20 & CP & 0.957 & 0.986 & 0.908 & \textbf{0.985} & 0.969 & \textbf{0.997} & 0.922 & \textbf{0.992} \\
 & EL & 0.164 & 0.382 & 0.165 & 0.247 & 0.196 & 0.533 & 0.197 & 0.326 \\ \hline

20,5,5,4,4 & CP & 0.962 & 0.985 & 0.910 & \textbf{0.983} & 0.968 & \textbf{0.998} & 0.921 & \textbf{0.992} \\
 & EL & 0.164 & 0.379 & 0.165 & 0.246 & 0.195 & 0.535 & 0.198 & 0.327 \\ \hline

7,7,7,7,7 & CP & 0.958 & 0.942 & 0.922 & 0.953 & 0.969 & 0.954 & 0.929 & 0.964 \\
 & EL & 0.173 & 0.298 & 0.171 & 0.214 & 0.206 & 0.389 & 0.206 & 0.269 \\ \hline

7,7,8,8,9 & CP & 0.958 & 0.942 & 0.925 & 0.954 & 0.968 & 0.956 & 0.936 & 0.964 \\
 & EL & 0.162 & 0.257 & 0.160 & 0.191 & 0.193 & 0.329 & 0.192 & 0.237 \\ \hline

\end{tabular}
\end{center}
\end{table}

\begin{table}
\begin{center}
\caption{  Empirical coverage probabilities and expected lengths of two-sided confidence intervals for the parameter of common CV under Weibull distribution for $k=10$.}\label{tab.sim6}

\begin{tabular}{|c|c|cccc|cccc|} \hline
 &  & \multicolumn{4}{|c|}{$\tau =0.1$} & \multicolumn{4}{|c|}{$\tau =0.2$} \\ \hline
$n_1,\dots ,n_{10}$ &  & MSLR & GV1 & GV2 & GV3 & MSLR & GV1 & GV2 & GV3 \\ \hline

4,4,4,4,4,4,4,4,4,4 & CP & 0.890 & 0.859 & 0.668 & 0.964 & 0.923 & \textbf{0.978} & 0.737 & \textbf{0.997} \\
 & EL & 0.053 & 0.136 & 0.050 & 0.083 & 0.109 & 0.408 & 0.104 & 0.232 \\ \hline

4,4,5,5,6,4,4,5,5,6 & CP & 0.886 & 0.867 & 0.689 & 0.952 & 0.927 & 0.893 & 0.768 & \textbf{0.972} \\
 & EL & 0.047 & 0.095 & 0.044 & 0.063 & 0.096 & 0.256 & 0.093 & 0.157 \\ \hline

6,5,5,4,4,6,5,5,4,4 & CP & 0.882 & 0.862 & 0.691 & 0.952 & 0.929 & 0.895 & 0.767 & 0.972 \\
 & EL & 0.047 & 0.095 & 0.045 & 0.063 & 0.096 & 0.255 & 0.093 & 0.157 \\ \hline

5,5,5,5,10,5,5,5,5,10 & CP & 0.882 & 0.869 & 0.731 & 0.937 & 0.926 & 0.872 & 0.802 & 0.955 \\
 & EL & 0.041 & 0.069 & 0.039 & 0.049 & 0.084 & 0.169 & 0.082 & 0.114 \\ \hline

10,5,5,5,5,10,5,5,5,5 & CP & 0.882 & 0.875 & 0.725 & 0.939 & 0.932 & 0.876 & 0.800 & 0.954 \\
 & EL & 0.041 & 0.068 & 0.039 & 0.049 & 0.083 & 0.169 & 0.082 & 0.114 \\ \hline

4,4,5,5,20,4,4,5,5,20 & CP & 0.877 & 0.875 & 0.733 & 0.937 & 0.930 & 0.914 & 0.808 & 0.968 \\
 & EL & 0.035 & 0.059 & 0.035 & 0.043 & 0.072 & 0.150 & 0.073 & 0.101 \\ \hline

20,5,5,4,4,20,5,5,4,4 & CP & 0.878 & 0.877 & 0.740 & 0.935 & 0.925 & 0.914 & 0.804 & 0.968 \\
 & EL & 0.035 & 0.058 & 0.035 & 0.043 & 0.072 & 0.150 & 0.073 & 0.101 \\ \hline

7,7,7,7,7,7,7,7,7,7 & CP & 0.881 & 0.881 & 0.748 & 0.928 & 0.925 & 0.879 & 0.827 & 0.947 \\
 & EL & 0.037 & 0.054 & 0.036 & 0.041 & 0.076 & 0.122 & 0.074 & 0.090 \\ \hline

7,7,8,8,9,7,7,8,8,9 & CP & 0.876 & 0.878 & 0.766 & 0.920 & 0.928 & 0.879 & 0.842 & 0.941 \\
 & EL & 0.035 & 0.048 & 0.033 & 0.038 & 0.071 & 0.107 & 0.070 & 0.082 \\ \hline

 &  & \multicolumn{4}{|c|}{$\tau =0.3$} & \multicolumn{4}{|c|}{$\tau =0.35$} \\ \hline

4,4,4,4,4,4,4,4,4,4 & CP & 0.957 & 1.000 & 0.781 & 1.000 & 0.968 & \textbf{1.000} & 0.808 & \textbf{1.000} \\
 & EL & 0.167 & 1.228 & 0.163 & 0.627 & 0.197 & 1.864 & 0.195 & 0.941 \\ \hline

4,4,5,5,6,4,4,5,5,6 & CP & 0.961 & 0.999 & 0.821 & 1.000 & 0.970 & \textbf{1.000} & 0.838 & \textbf{1.000} \\
 & EL & 0.149 & 0.614 & 0.146 & 0.340 & 0.176 & 0.978 & 0.175 & 0.512 \\ \hline

6,5,5,4,4,6,5,5,4,4 & CP & 0.959 & 0.998 & 0.819 & 0.998 & 0.967 & \textbf{1.000} & 0.841 & \textbf{1.000} \\
 & EL & 0.149 & 0.620 & 0.147 & 0.343 & 0.176 & 0.971 & 0.175 & 0.509 \\ \hline

5,5,5,5,10,5,5,5,5,10 & CP & 0.960 & 0.946 & 0.849 & 0.984 & 0.968 & \textbf{0.995} & 0.872 & \textbf{0.996} \\
 & EL & 0.129 & 0.350 & 0.128 & 0.216 & 0.154 & 0.509 & 0.153 & 0.297 \\ \hline

10,5,5,5,5,10,5,5,5,5 & CP & 0.959 & 0.949 & 0.855 & 0.985 & 0.970 & \textbf{0.995} & 0.875 & \textbf{0.997} \\
 & EL & 0.129 & 0.349 & 0.128 & 0.216 & 0.154 & 0.509 & 0.153 & 0.297 \\ \hline

4,4,5,5,20,4,4,5,5,20 & CP & 0.958 & 0.992 & 0.853 & 0.993 & 0.967 & \textbf{0.999} & 0.869 & \textbf{0.998} \\
 & EL & 0.112 & 0.344 & 0.113 & 0.202 & 0.133 & 0.523 & 0.136 & 0.286 \\ \hline

20,5,5,4,4,20,5,5,4,4 & CP & 0.957 & 0.993 & 0.853 & 0.994 & 0.967 & \textbf{1.000} & 0.869 & \textbf{0.997} \\
 & EL & 0.112 & 0.345 & 0.114 & 0.203 & 0.133 & 0.524 & 0.135 & 0.287 \\ \hline

7,7,7,7,7,7,7,7,7,7 & CP & 0.960 & 0.891 & 0.876 & 0.954 & 0.968 & 0.907 & 0.892 & \textbf{0.966} \\
 & EL & 0.118 & 0.223 & 0.116 & 0.154 & 0.140 & 0.296 & 0.139 & 0.196 \\ \hline

7,7,8,8,9,7,7,8,8,9 & CP & 0.946 & 0.894 & 0.890 & 0.955 & 0.965 & 0.906 & 0.910 & 0.960 \\
 & EL & 0.109 & 0.189 & 0.109 & 0.136 & 0.131 & 0.248 & 0.131 & 0.171 \\ \hline

\end{tabular}
\end{center}
\end{table}

\section{Real examples}

\label{sec.ex}

\begin{example}
 In this part, we used the data set given by
 \cite{fu-ts-98}.
{\  This data set is also analyzed by
 \cite{ja-ka-13} and
 \cite{kr-le-14}
 for the problem of testing the equality of several normal independent CV's, and considered by
 \cite{tian-05} and
 \cite{be-ja-08}}
 for the problem of inference about the common CV. The Hong Kong Medical Technology Association has conducted a Quality Assurance Programme for medical laboratories since 1989 with the purpose of promoting the quality and standards of medical laboratory technology. The data are collected from the third surveys of 1995 and 1996 for the measurement of Hb, RBC, MCV, Hct, WBC, and Platelet in two blood samples (normal and abnormal).  The summary statistics for this subset of data is given in Table \ref{tab.ex1}. The main data set of this study has  not been presented, and therefore, we cannot check the normality assumption.

At level $\alpha =0.05$,
 \cite{ja-ka-13}
  showed that the CV for RBC, MCV, Hct, WBC, and Plt in 1995 is not significantly different from that of 1996 in the abnormal blood samples.  The confidence intervals for the common CV based on our proposed MSLR method and the three generalized approaches for these data between 1995 and 1996 in each measurement are given in Table \ref{tab.ex2}. Since the sample sizes are large, the results of all methods are close to each other.
\end{example}

\begin{table}[ht]
\begin{center}
\caption{ Summary statistics of measurements in the abnormal blood samples.}\label{tab.ex1}

\begin{tabular}{|c|c|ccccc|} \hline
Year &  & RBC & MCV & Hct & WBC & Plt \\ \hline

1995 & $n_1$ & 65 & 63 & 64 & 65 & 64 \\
 & ${\bar{x}}_1$ & 4.606 & 87.25 & 0.4024 & 17.68 & 524.7 \\
 & $s_1$ & 0.0954 & 3.496 & 0.0194 & 1.067 & 37.05 \\ \hline

1996 & $n_2$ & 73 & 72 & 72 & 73 & 71 \\
 & ${\bar{x}}_2$ & 4.574 & 92.33 & 0.4216 & 18.93 & 466.5 \\
 & $s_2$ & 0.0838 & 3.078 & 0.0168 & 1.211 & 41.58 \\ \hline
\end{tabular}

\end{center}
\end{table}

\begin{table}[ht]
\begin{center}
\caption{The two-sided confidence intervals for the common CV of measurements in the abnormal blood samples.}\label{tab.ex2}

\begin{tabular}{|c|ccccc|} \hline
Method & RBC & MCV & Hct & WBC & Plt \\ \hline

MSLR & (0.017,0.022) & (0.033,0.042) & (0.039,0.050) & (0.056,0.071) & (0.072,0.092) \\
GV1  & (0.017,0.022) & (0.034,0.042) & (0.039,0.050) & (0.056,0.071) & (0.072,0.092) \\
GV2  & (0.017,0.022) & (0.033,0.041) & (0.039,0.049) & (0.056,0.071) & (0.071,0.090) \\
GV3  & (0.017,0.022) & (0.034,0.041) & (0.039,0.050) & (0.056,0.071) & (0.072,0.091) \\ \hline
\end{tabular}
\end{center}
\end{table}

\begin{example}
 The data set in Appendix D of
 \cite{fl-ha-91}
 refer to survival times of patients from four hospitals. It is analyzed by \cite{na-ra-03} and \cite{be-ja-08}.  These data and their descriptive statistics are given in Table \ref{tab.ex3}.
 The normality assumption for survival times of patients in each of the hospitals was checked using   Kolmogorov-Smirnov (KS) and Shapiro-Wilk (SW) tests.   The p-values are given in Table \ref{tab.ex3}.
 Therefore, the normal model appears to be appropriate for each group.

 \cite{na-ra-03}
 tested homogeneity of CV's for the hospitals and they showed that all tests give the same conclusion of accepting the null hypothesis. Therefore, we have a common CV for these data. The two-sided confidence intervals for the common CV based on MSLR, GV1, GV2 and GV3 are (0.4748, 0.5988), (-1.7855, 3.6561), (0.4568, 1.1759) and (-0.5457, 2.2563), respectively. It easily can be seen that the lengths of these methods are 0.1240, 5.4416, 0.7191, and 2.8020, respectively. Therefore, the length of the confidence interval proposed by
 \cite{tian-05}
 is larger than other methods while the length of our proposed confidence interval is smaller than other methods. This is consistent with the simulation results in Section \ref{sec.infer} that  the length of our proposed method is smaller than other approaches.
\end{example}

\begin{table}
\begin{center}
\caption{ Data and descriptive statistics for survival times of patients from four hospitals.}\label{tab.ex3}

\begin{tabular}{|c|l|rr|cc|} \hline
 & Data & ${\bar{x}}_i$ & $s^2_i$                     & {\  KS} & {\ SW}\\ \hline
Hospital  1 & 176  105  266  227  66 & 168.0 & 6880.5 &0.990 &0.794 \\
Hospital  2 & 24  5  155  54 & 59.5 & 4460.3          &0.822 &0.309 \\
Hospital  3 & 58  64  15 & 45.7 & 714.3               &0.748 &0.215 \\
Hospital  4 & 174 42  305  92 30  82  265  237 208  147 & 154.6 & 8894.7 &0.939& 0.695 \\ \hline
\end{tabular}

\end{center}
\end{table}

%

\section{Conclusion}
\label{sec.con}
In this paper, we utilize the method of modified signed log-likelihood ratio for the inference about the parameter of common coefficient of variation in several independent normal populations. Also, we compared it with other competing approaches known as generalized variable approaches in terms of empirical coverage probabilities and expected lengths. Simulation studies showed that
the coverage probability of the MSLR method  is  close to the confidence coefficient and
its expected length is shorter than expected lengths of the GV methods. Therefore, our proposed approach
acts very  satisfactory regardless of the number of samples and for all different values of common CV, even for small sample sizes, while the generalized variable approaches  act well when the value of common CV is large.  It is notable that an executable program written in R is provided to compute the confidence intervals for the common CV  and can be made available to any interested reader.

\bibliographystyle{apa}

\end{document}